\documentclass{bmcart}

\usepackage{amsthm,amsmath}
\RequirePackage{natbib}
\RequirePackage{hyperref}
\usepackage[utf8]{inputenc} 

\usepackage{graphicx}
\usepackage{caption}
\usepackage{subcaption}

\startlocaldefs

\endlocaldefs

\begin{document}

\begin{frontmatter}

\begin{fmbox}
\dochead{Research}


\title{A large scale prediction of bacteriocin gene blocks suggests a wide functional spectrum for bacteriocins}


\author[
   addressref={aff1},                   
   email={jamietmorton@gmail.com}   
]{\inits{JTM}\fnm{James T} \snm{Morton}}
\author[
   addressref={aff2,aff3},
   email={stefan_freed@nd.edu}
]{\inits{SDF}\fnm{Stefan D} \snm{Freed}}
\author[
   addressref={aff2},
   email={shaun_lee@nd.edu}
]{\inits{SWL}\fnm{Shaun W} \snm{Lee}}
\author[
   addressref={aff1,aff4, aff5},  
   noteref={n1},
   corref={aff5},
   email={idoerg@iastate.edu}
]{\inits{IF}\fnm{Iddo} \snm{Friedberg}}


\address[id=aff1]{
  \orgname{Department of Computer Science and Software engineering}, 
  \street{Miami University},                      
  \city{Oxford, OH},                              
  \cny{USA}                                    
}

\address[id=aff2]{%
  \orgname{Eck Institute for Global Health, Department of Biological Sciences,},
  \street{University of Notre Dame},
  \city{South Bend, IN},
  \cny{USA}
}

\address[id=aff3]{%
  \orgname{Chemistry Biochemistry Biology Interface Program},
  \street{University of Notre Dame},
  \city{South Bend, IN},
  \cny{USA}
}

\address[id=aff4]{%
  \orgname{Department of Microbiology},
  \street{Miami University},
  \city{Oxford, OH},
  \cny{USA}
}

\address[id=aff5]{%
  \orgname{Department of Veterinary Microbiology and Preventive Medicine},
  \street{Iowa State University},
  \city{Ames, IA},
  \cny{USA}
}


\begin{artnotes}
\note[id=n1]{Corresponding Author} 


\end{artnotes}

\end{fmbox}


\begin{abstractbox}

\begin{abstract} 
Bacteriocins are peptide-derived molecules produced by bacteria, whose recently-discovered functions include virulence factors and signaling molecules as well as their better known roles as antibiotics. To date, close to five hundred bacteriocins have been identified and classified.  Recent discoveries have shown that bacteriocins are highly diverse and widely distributed among bacterial species. Given the heterogeneity of bacteriocin compounds, many tools struggle with identifying novel bacteriocins due to their vast sequence and structural diversity. Many bacteriocins undergo post-translational processing or modifications necessary for the biosynthesis of the final mature form. Enzymatic modification of bacteriocins as well as their export is achieved by proteins whose genes are often located in a discrete gene cluster proximal to the bacteriocin precursor gene, referred to as \textit{context genes} in this study. Although bacteriocins themselves are structurally diverse, context genes have been shown to be largely conserved across unrelated species. Using this knowledge, we set out to identify new candidates for context genes which  may clarify how bacteriocins are synthesized, and identify new candidates for bacteriocins that bear no sequence similarity to known toxins. To achieve these goals, we have developed a software tool, \textbf{B}acteriocin \textbf{O}peron and gene block \textbf{A}ssociator \textbf{(BOA)} that can identify homologous bacteriocin associated gene clusters and predict novel ones. We discover that several phyla have a strong preference for bactericon genes, suggesting distinct functions for this group of molecules.
Availability: https://github.com/idoerg/BOA
\end{abstract}


\begin{keyword}
\kwd{bacteriocins}
\kwd{operons}
\kwd{microbiology}
\kwd{gene blocks}
\end{keyword}


\end{abstractbox}
%

\end{frontmatter}



\section*{Background}
 
Natural Product discovery has been a cornerstone of many pharmaceuticals and therapeutics. It is estimated that about 80\% of all drugs are either natural products or derived analogs\cite{who1985plants}. These compounds encompass antibiotics (penicillin, tetracycline, erythromycin), anti-infectives (avermectin, quinine, artemisinin), pharmaceuticals (lovastatin, cyclosporine, rapamycin) and anticancer drugs (taxol, doxorubicin) \cite{Harvey2008drug}. Yet, despite this long history of success, pharmaceutical efforts in natural products research has decreased steadily between 2001 and 2008 \cite{li2009naturalproducts}. Financial pressure from drug companies as well as difficulties in isolation and identification of natural compounds have severely limited the discovery rate of these important sources.

Bacteriocins are a large class of peptide-based antibiotics that encompass an extraordinary amount of chemical, structural, and functional diversity\cite{Willey2007lantibiotics}. 
The structures of these compounds have revealed that a large number comprise a class of highly modified polypeptides. Interestingly, many bacteriocins are synthesized ribosomally as precursor peptides and are subsequently modified post-translationally to yield their biologically active forms. Post-translational modification serves to confer specific chemical properties that could not be obtained by peptide synthesis alone. Furthermore, post-translational modifications can be used as a mechanism to control the activation of the toxic activities of the bacteriocin, and thus exert a level of control and host immunity\cite{Guder2000lantibiotics}.

Genome mining has been an important technological resource in the discovery of novel natural products, including bacteriocins. Bacteriocin-like peptides are highly attractive candidates for genome mining, as these natural products are genetically encoded with nearby genes encoding their corresponding modifying enzymes. Proximity to genes encoding known modifying enzymes can often aid in the identification of  peptide biosynthesis gene clusters \cite{arnison2013ribopepnomenclature}. In many cases, several metabolites have been identified from ``cryptic'' or ``orphan gene clusters'' \cite{perez2014bacteriocins}. These cryptic gene clusters have demonstrated that new, as yet uncharacterized enzymology is likely to be involved in the assembly of the final natural product, likely leading to a greater diversity of bacterial peptides than previously appreciated. 

Several web-based gene mining and annotation tools have been developed to aid in the identification, characterization, and classification of novel bacteriocins. These include mining tools such as BAGEL\cite{Heel2013bagel3} and bacteriocin repositories such as BACTIBASE\cite{Hammami2007bactibase}. Anti-SMASH is a recently developed website that expands genome mining to not only bacteriocins but a host of other genetically-identifiable antibiotics and other secondary metabolites\cite{Blin2013antismash} Although gene mining resources such as BAGEL have existed for many years, one major limitation in bacteriocin gene mining that is becoming evident as more bacteriocins are identified is the lack of homology in the genes encoding the actual precursor peptides for bacteriocins. Detecting novel bacteriocin producing genes through bioinformatic methods focused on precursor peptide discovery therefore remains a significant challenge.  

The bacteriocins' structure and sequence diversity make them difficult to detect using sequence homology algorithms such as BLAST\cite{Altschul1997Gapped}.  In addition, their short sequence length makes them difficult to detect with ORF calling tools. Programs such as BAGEL have made great strides in identifying bacteriocins by searching for well-defined conserved motifs within the bacteriocin toxin sequences and adjacent context genes using Pfam databases\cite{Heel2013bagel3}. However, the number of hypothetical bacteriocin genes that can be identified using such a method is highly limiting as these motifs are not necessarily known or conserved. Given that bacteriocin genes are highly diverse, it is likely that mining genomes for potential bacteriocin genes using sequence or profile similarity are likely to miss a large number of as-yet identified bacteriocin compounds. 

Genes that encode bacteriocin precursors are often proximal in genomic sequence to accessory genes that are required to modify and secrete the bacteriocin peptide. These co-located \textit{gene blocks} often contain genes that encode enzymes which perform post-translational modification and maturation of the bacteriocin product. Additionally, many of these gene blocks contain genes encoding various transporter proteins, presumably linked to specific export of the mature bacteriocin. Unlike the bacteriocin genes, these \textit{context} genes are conserved across taxa, as they belong to conserved families of enzymes and other modifying proteins, as well as to transporters such as ATP-binding cassette (ABC) family of transporters. In the case of the large family of thiazole and oxazole modified bacteriocins, such as Microcin~B17 and Streptolysin~S, each gene block shares both conserved modification proteins as well as transport machinery genes that are located close to the bacteriocin gene \cite{Lee2008Discovery,arnison2013ribopepnomenclature}.  

Here we present a new methodology that takes advantage of the conservation of context genes to identify locations of gene blocks associated with bacteriocins. Our approach is to identify context genes in addition to toxin genes without restricting our tool to finding homologs in annotated databases. To the best of our knowledge, there is no available public database that contains the complete bacteriocin gene clusters and their associated context genes. Therefore, BOA is useful for the construction of such a tool as well as for mining genomes for putative bacteriocins. We provide bacteriocin gene block predictions for 2773 genomes, in which the bacteriocin gene blocks may be browsed. The method is implemented as a software tool named \textbf{B}acteriocin \textbf{O}peron and gene block \textbf{A}ssociator \textbf{(BOA)}

\section*{Methods}
We used the bacterial and archaeal genome files from GenBank 2014\cite{Benson2014GenBank} for our dataset.  Since we are interested in identifying bacteriocin associated gene clusters, we did not analyze partial contig files, and only whole bacterial chromosomes and plasmids were analyzed. Bacteriocins identified by BAGEL3 and seven experimentally identified bacteriocin associated gene blocks were used as a standard of truth or a ``gold standard'' data set. All of the context genes within these gene blocks were placed into five functional categories \- toxins, modifiers, immunity, transport, and regulation. Toxin genes refer to the open reading frames encoding the toxin precursor peptide;  modifiers perform post-translational modification to the protoxin, and can include enzymes which are involved in amino acid modification (cyclodehydration, lanthionine synthesis, as well as leader peptide processing enzymes);  immunity genes that prevent the toxin from affecting the host bacterial cells;  transport genes create transporter proteins to move the toxin outside of the cell, and  regulator genes control the expression of toxin proteins and other genes in the operon \cite{arnison2013ribopepnomenclature}. 



An overview of the pipeline is shown in Figure~\ref{fig:pipeline}. The method in detail is as follows:
\begin{enumerate}
\item Construct a set of experimentally-verified context genes. This we call the Literature Curated Set. The LC Set includes seven known bacteriocin associated gene blocks. The known gene blocks comprised enterococcal cytolysin and enterocin AS-48 from \textit{Enterococcus faecalis}, microcin J25 from \textit{Escherichia coli}, Nisin A from \textit{Lactococcus lactis} subsp. \textit{lactis}, streptolysin S and salivaricin A from \textit{Streptococcus pyogenes}, and thiocillin from \textit{Bacillus cereus} \cite{cobourn2003cytolysin,galvez1989as-48,salomon1992microcinJ25,kaletta1989nisin,
nizet2000sls,ross1993salivaricin,shoji1976isolation}.  These blocks are representative of several important classes of bacteriocins, namely lantibiotics (enterococcal cytolysin, Nisin A, salivaricin A), thiopeptides (thiocillin), thiazole/oxazole-modified microcins (TOMMs; streptolysin S), lassoed tail peptides (microcin J25), and circular bacteriocins (enterocin AS-48).  While bacteriocin biosynthetic gene blocks are widely distributed among prokarya, the major structurally-related groups are few and their relevant context genes are largely represented in the LC Set. The genes were categorized as toxins, modifiers, immunity, transport, or regulation.
The genes making up the LC Set are available in the Supplementary Material.
\item BLAST the LC Set genes and the BAGEL toxin genes against the bacterial genome set. 
\item Select ORFs with (a)  e-value $<10^{-5}$ and (b) within $\pm$ 50kb of the homologs to the toxin genes (whether from the LC set or the BAGEL set).
\item Assign each homologous ORF according to the category of the gene to which it is found to be similar: toxin, modifier, immunity, transport, or regulation. Ambiguities are resolved by  taking the best hit based on the BLAST score.
\item Build profile HMMs: cluster the sequences in each bin independently using CD-HIT \cite{Li2006cdhit}, then use MAFFT~\cite{Katoh2013mafft} to perform a multiple alignment in each homology cluster, and HMMER ~\cite{Eddy2011Accelerated} to build pHMMs from the multiple alignments. 
\item Run the resulting pHMM's against the bacterial genome files. Since most of the HMMER hits are probably false positives, we set a score threshold to filter them out. We determined this threshold by obtaining HMMER scores for all of the BAGEL bacteriocins found on the bacterial genomes and choosing the lowest BAGEL score as the threshold. See Figure~\ref{fig:scores}
\item Use a clique filter (see below) to identify those genes that are close together and therefore candidates for bacteriocin gene blocks. See Figure~\ref{fig:clique}.
\end{enumerate}


To identify gene blocks that are candidates for bacteriocin biosynthesis, we used a \textit{clique filter}. A clique is a complete subgraph where any two nodes are connected by an edge.  We created graphs from the $\pm$50kb regions where genes are represented by nodes and for every pair of genes that are within 25kb of each other, an edge is created between them. The detected cliques are estimated gene blocks such that all of the genes are within 25kb of each other. This 25kb threshold was based on the size of known bacteriocin gene clusters given in our gold standard data set. Figure~\ref{fig:clique} illustrates the use of the clique filter to identify potential bacteriocin gene blocks.

A major challenge in finding toxin genes is that due to the short length and low complexity of these peptides, many genes will be missed because of ORF calling errors or lack of similarity to known toxin genes. To overcome this problem, we organized all detected cliques into gene blocks with homologs to known toxin genes and gene blocks without known toxin genes. Cliques with known toxin genes are required to have at least one toxin gene and one transport gene. Cliques without any known toxin genes are required to have at least one of each of a modifier, transport, immunity, and regulator genes. In this way, we ensure that the cliques we identify have the needed components to present a putative bacteriocin biosynthetic locus, and the toxin gene can later be searched using less restrictive procedures.

\section*{Results}
The many characterized bacteriocins have seldom been experimentally validated in parallel in the multiple species which putatively code for their production, restricting our standard-of-truth data set to a small group of well-studied bacteriocins relative to the large number of organisms that produce them.  In addition, estimating the false positive rate is difficult, requiring excessive \textit{de novo} experimental validation. Therefore, to evaluate BOA's performance, we compared the bacteriocins that we have found to BAGEL bacteriocins found in bacterial genomes.

To compare BOA against BAGEL, the toxins shared between BOA and BAGEL were identified using BLAST using the default parameters.  
As shown in Table~\ref{tab:detected}, BOA only missed 22 (5\%) bacteriocins that BAGEL detected, while predicting 457 (95\%) in agreement with BAGEL.  In addition, BOA was able to predict over 1003 more putative bacteriocins on bacterial genomes than BAGEL. In addition, BOA identified 83 regions that are highly likely to be associated with bacteriocin production.



The detected gene blocks were classified into five groups: (1) gene blocks with all five functional classes (toxin, modifier, immunity, regulator, transport), (2) gene blocks with only four functions, (3) three functions, (4) two functions, and (5) unknown toxins. Figure~\ref{fig:function_count} shows these findings. 

Previously it has been established that every bacteriocin locus needs at minimum a toxin gene and an immunity gene\cite{dimov2005lacticacid}. The gene blocks in the first four groups in Figure~\ref{fig:operon_cnts} and Figure~\ref{fig:operon_ratio} all have at least one toxin and one transport gene. The final group of gene blocks do not have any identified bacteriocin genes, but each detected gene block is required to have all of the other genes. This final group contains likely candidates for bacteriocin-associated gene blocks that do not yet have a known, identified bacteriocin.  From these findings, it is evident that genes categorized as transport genes were identified to be the most common type of context gene. 

Interestingly, the three species harboring the greatest number of predicted bacteriocin-associated gene blocks in our screen inhabit different ecological niches.  \textit{Streptococcus equi subsp. zooepidemicus} is a common colonizer of the respiratory tract with the capacity for opportunistic infection in a variety of domesticated animals and sometimes severe infections in humans following zoonotic transmission \cite{Minces2011Human}.  \textit{Streptomyces griseus} is a soil-dwelling bacterium that has been studied and utilized in the biotechnology industry for production of numerous secondary metabolites including the first aminoglycoside antibiotic, streptomycin \cite{ohnishi2008Genome}.  Finally, \textit{Leifsonia xyli} is a pathogenic obligate colonizer of the xylem of host plants, causing economically damaging ratoon stunting disease in sugarcane \cite{Ghai2014Rapid}.  The radically different environments in which these bacteria reside suggest that predicted bacteriocins must have distinct functions, specific organism targets, or both.  Likewise, functional validation and characterization of these kinds of predicted bacteriocins must take into account the niche in which its producing organism resides, probing functionalities that target ecologically relevant target organisms.  

Among the gene blocks identified by BOA was the recently described and experimentally characterized caynothecamide biosynthetic locus of \textit{Cyanothece sp. PCC 7425} (GenBank: CP001344.1) \cite{donia2011cyanobactinmining}.  This gene block, part of the patellamide family, has nine predicted precursor peptide ORFs with conserved N-termini and divergent C-termini likely resulting from repeated precursor duplication and divergence.  Other bacteriocin clusters have been described with only one precursor peptide duplication, and such examples may represent an early step toward the substrate elaboration displayed by the cyanothecamides \cite{Tabata2013Novel}.  Interestingly, most of the cyanothecamide putative precursors lack the canonical pentapeptide motif required for patellamide maturation, suggesting that inclusion of these sequences in the BOA gold standard set could expand the set of identified putative toxin genes to include other non-canonical substrates.  Only two of the nine cyanothecamide putative toxin genes have been experimentally implicated as precursors to identifiable mature patellamide-like compounds \cite{houssen2012cyanothecamides}. Yet, the capacity for biosynthetic machinery to modify substrate peptides with suitable N-terminal domains despite drastic variability in the C-terminal portion of the peptide has been demonstrated in other bacteriocins \cite{mitchell2009strucfuncsls,haft2010expandedtomms}.  The features of this particular gene block raise the possibility that bacteriocin loci encoding post-translationally modified peptides could, through elaboration of sequence diversity in multiple cognate peptide substrates, confer a greater breadth of functional diversity to producing organisms than previously appreciated \cite{sardar2015cyanobactinevolution}.

Within the genome of the important human pathogen Group A \textit{Streptococcus} from which two members of our gold standard set were obtained (Streptolysin S and Salivaricin A), BOA also identified the gallidermin-related lantibiotic Streptin \cite{karaya2001streptindiscovery}.  Despite the experimental validation of Streptin as an active bacteriocin, little further insight has been gained into the role of Streptin with respect to pathogenic infection or colonization dynamics \cite{wescombe2003streptincharacterization}. Identification and subsequent experimental validation of bacteriocins in important human pathogens like Group A \textit{Streptococcus} will likely yield insights into the biology and biochemistry of pathogenic colonization, especially given the current explosion of interest in the human microbiome and probiotic disease interventions.

Of the 1054 species with identified bacteriocins, only 11 species were from the domain Archaea out of 360 Archaea genomes in GenBank. From the 1043 bacterial species with bacteriocins identified, the majority of them are identified as either Proteobacteria or Firmicutes. The exact breakdown of the phyla and their corresponding mean function counts is shown in Table~\ref{tab:phyla_cnts}. It is important to note that our finding does not imply that most bacteriocin producing bacteria are Firmicutes and Proteobacteria, or that bacteriocins are rare in Archaea. It is more likely that previous research in identifying bacteriocins was biased towards the former two phyla.

From Table~\ref{tab:phyla_cnts}, it is apparent that mean distribution of gene types between phyla are very different.  For instance, Firmicutes have significantly more immunity genes than any of the other phyla.  Also, gene blocks found in Firmicutes, Actinobacteria and Cyanobacteria have a higher toxin gene count than other bacterial phyla.

\section*{Conclusions}
To our knowledge, BOA is the first time a curated data set has been established for bacteriocin context genes. Even with seven different bacteriocin gene blocks as a gold standard set, our method has identified several hundred putative bacteriocin gene blocks, most of which have not been previously annotated.  We believe that even more homologous gene blocks can be identified with a larger validated database of context genes.  Additionally, upon manual inspection of some predicted blocks, some nearby putative ORFs appeared likely to be involved in predicted bacteriocin biosynthesis but were not identified by BOA.  This may permit a manually curated strategy whereby one may subjectively designate putative context genes from a BOA-predicted bacteriocin gene block and feed the more richly-annotated gene block back into BOA as a new member of the now-expanded gold standard set. Such an approach could serve to iteratively extend the phylogenetic boundaries of BOA in a controlled way each time the limits of similarity are reached.  We are currently exploring the merits of this approach. The widespread prevalence and diversity of bacteria having bacteriocins and their highly varied lifestyles suggest early ancestry and a subsequent adaptation of these gene blocks to the specific functional needs of the bacteria producing them. Previous studies have shown that  bacteriocin context genes tend to be in bacteria that share an environmental niche despite phylogenetic disparity, suggesting that functional adaptation is likely to be a major mechanism for bacteriocin design and production \cite{Lee2008Discovery}.

BOA was able to identify the majority of bacteriocin gene clusters that BAGEL identified. BOA also predicted over seven times more bacteriocins in whole bacterial genomes than BAGEL, including many identifiable bacteriocin gene blocks with experimental validation. Because BOA encompasses a large number of taxa, the information in BOA can also be used to explore the evolutionary development of bacteriocin gene blocks and how different biosynthetic loci have evolved in different clades. Finally, BOA has assembled the first dataset that contains information about homologous bacteriocin genes and their associated gene clusters.


\begin{backmatter}

\section*{Competing interests}
  The authors declare that they have no competing interests.

\section*{Author's contributions}
    IF, SWL and JTM conceived the idea. IF and JTM designed the experiment. JTM wrote the code and ran the analyses. SWL and SDF provided the gold-standard data and interpreted the results. All authors wrote the manuscript.

\section*{Acknowledgments}
  We are grateful to Sean Eddy and Rob Finn for the use of the HMMER logo. Some images used in Figure 1 are reproduced from Wikimedia Commons under CC-BY 3.0 or 4.0 license. We gratefully acknowledge the support of the Miami University High Performance Computing Facility. This work was supported, in part, by National Science Foundation grant ABI-1146960 (IF) and NIH 1DP2OD008468-01. (SWL). SDF was supported, in part by training grant NIH T32GM075762.
  

\bibliographystyle{bmc-mathphys} 
\bibliography{boa_ref}      



\clearpage
\section*{Figures}

\begin{figure}[h!]

\includegraphics[scale=0.45]{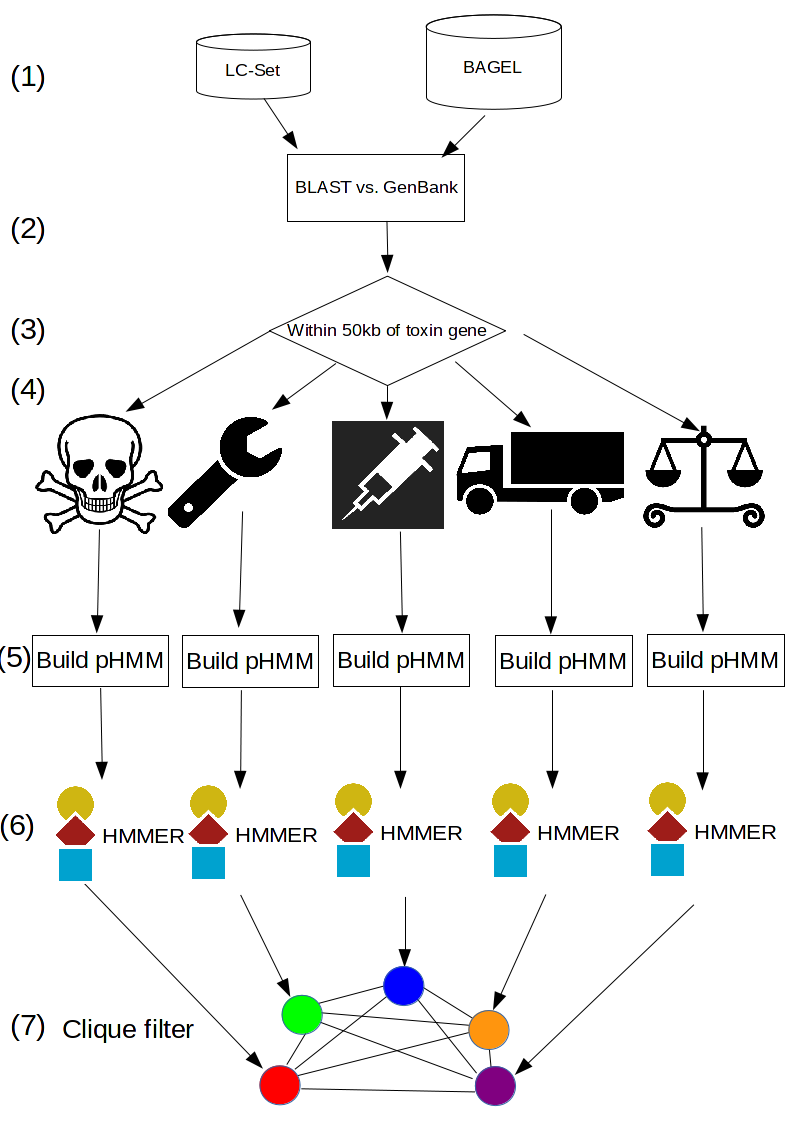}
\caption{\footnotesize{\csentence{An overview of the BOA Pipeline.} The stages in the pipeline are elaborated upon in the Methods section. (1): construction of the LC-Set and the BAGEL set; (2) BLAST LC and BAGEL genes against all bacterial \& archaeal genomes evalue=$10^{-5}$; (3) select the ORFs within $\pm 50$kb of homologs to toxin genes (4) assign ORFs to one of the following classes (left to right): toxin, modifier, immunity, transport, regulation; (5) build pHMMs from each category: cluster sequences using CD-HIT, align sequences in each cluster using MAFFT, then use hmmbuild from the HMMER suite to construct HMMs; (6) run hmmsearch from the HMMER suite against the genome files to extract more sequences from each category, remove predicted false positives using a threshold score as explained in Methods (7) use a clique filter to identify genes that are close together.}}	
\label{fig:pipeline}
\end{figure}
\clearpage
\begin{figure}[h!]

\includegraphics[scale=0.5]{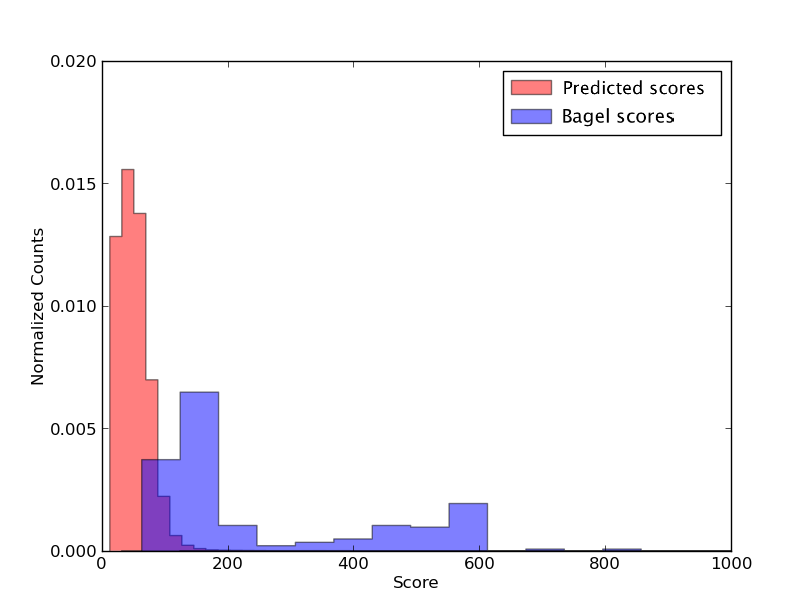}
\caption{\footnotesize{\csentence{Determining the threshold for similarity-based search of toxin genes.} Toxin gene candidates were derived as described in the text. To determine an adequate threshold for inferring homology, we examined the distribution of HMMER scores for homologs for predicted toxin genes (red) and BAGEL-derived toxin genes (blue). BAGEL toxin gene scores were used to set a minimum threshold of acceptance for HMMER scores for predicted genes.}}	
\label{fig:scores}
\end{figure}
\clearpage

\begin{figure}[h!]

\includegraphics[scale=0.2]{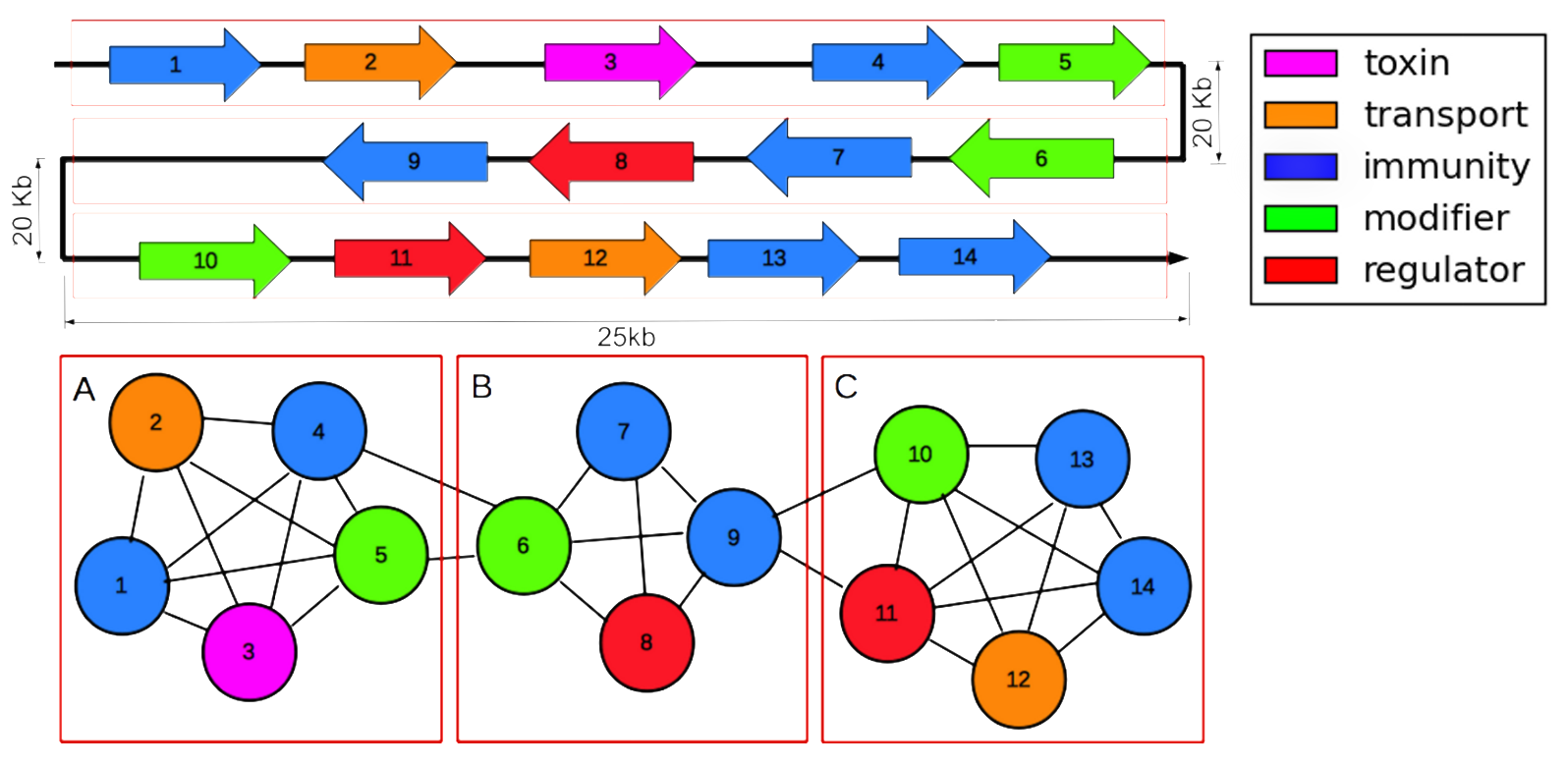}
\caption{\footnotesize{\csentence{Using a clique filter to identify putative bacteriocin gene blocks.} Drawing is not to scale, and other genes may exist in between those shown. A, B and C are all cliques formed from these genes.
Clique A has an identifiable homolog to a known toxin, and is considered a viable candidate for a toxin gene block.
Clique B does not have all necessary functions, and therefore is not considered to be a candidate.
Clique C contains all necessary functions and therefore is considered a candidate for a bacteriocin gene block or operon even though there is no homology-detected toxin gene.}}	
\label{fig:clique}
\end{figure}
\clearpage

\begin{figure}[h!]
\includegraphics[scale=1.0]{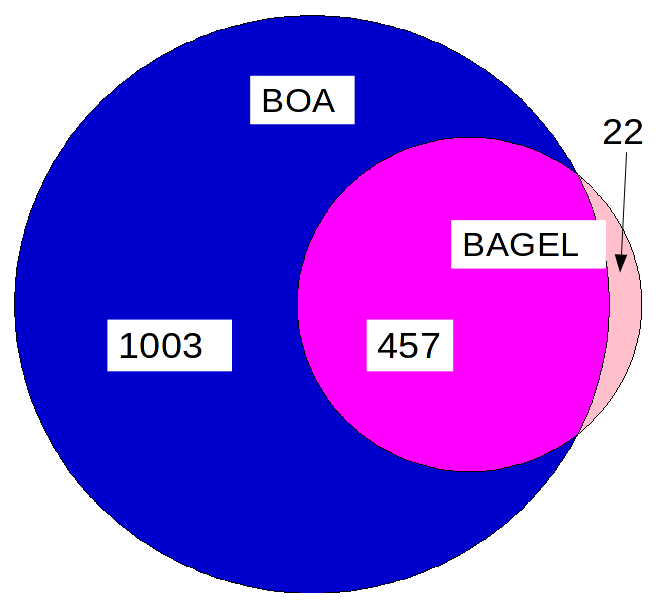}
\caption{\footnotesize{95\% (457 out of 479) of BAGEL toxins were predicted by BOA. BOA predicted an additional 1003 toxins throughout bacterial genomes that are not listed in BAGEL. Twenty-two BAGEL toxins were not predicted by BOA.}}	
\label{fig:operon_cnts}
\end{figure}

\begin{figure}[h!]
\begin{subfigure}[b]{0.5\textwidth}
  \includegraphics[width=\linewidth]{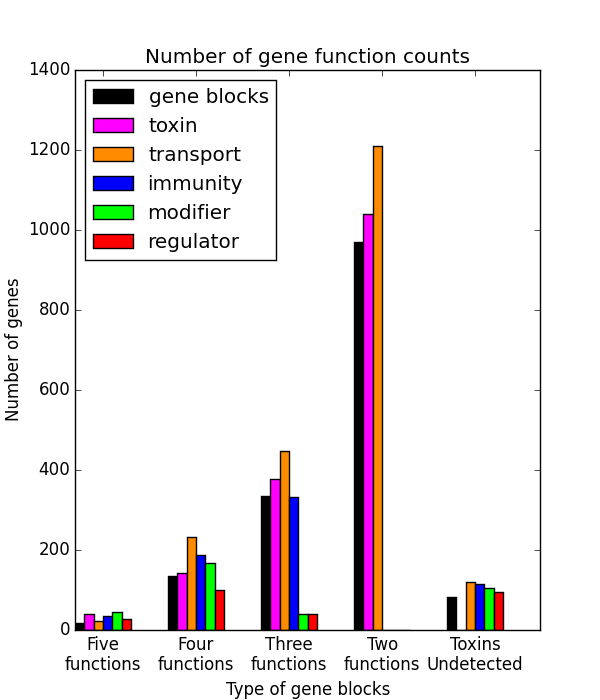}
  \caption{Gene counts}
\end{subfigure}
\begin{subfigure}[b]{0.5\textwidth}
  \includegraphics[width=\linewidth]{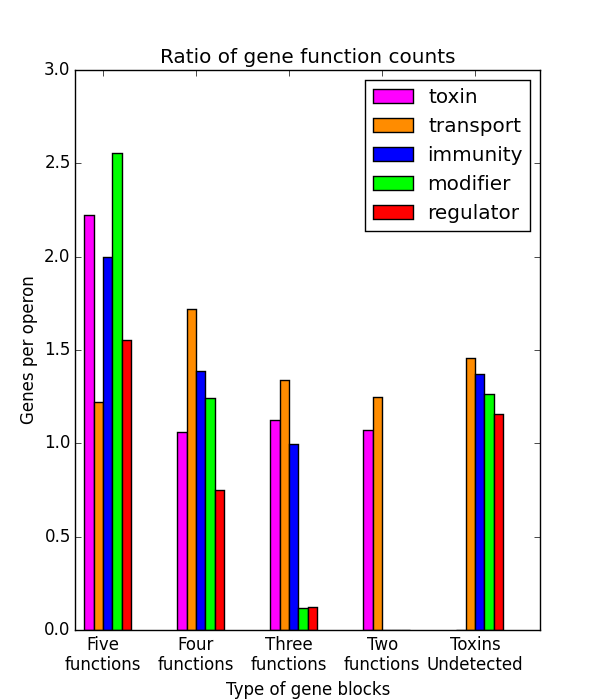}
  \caption{Gene counts per gene block}\label{fig:operon_ratio}
\end{subfigure}
\caption{\footnotesize{Gene blocks were classified by the number of detected functions. (a): number of total genes found; (b) gene counts per detected block.}}
\label{fig:function_count}
\end{figure}
\clearpage

\begin{figure}[h!]

\includegraphics[scale=0.10]{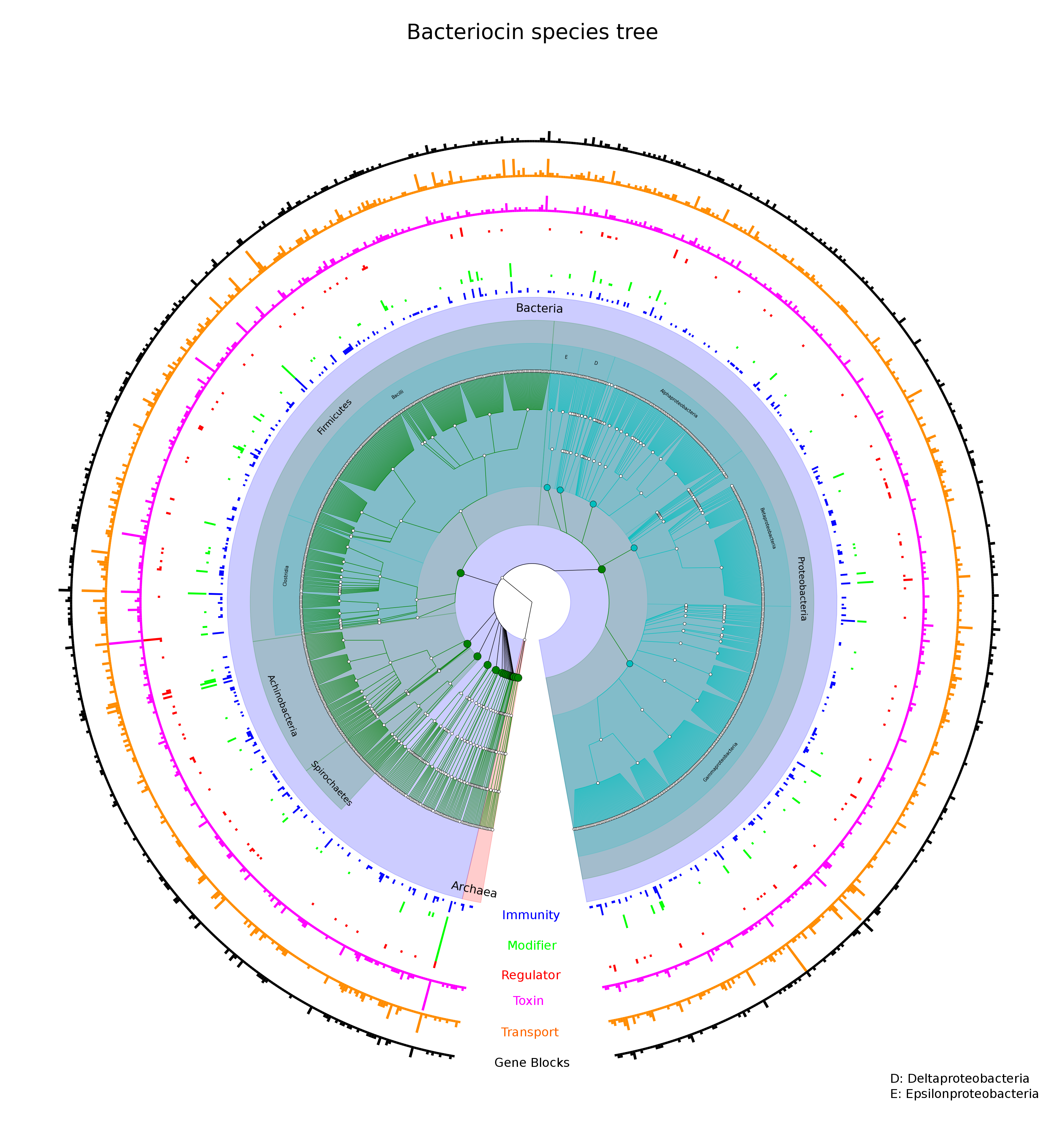}
\caption{\footnotesize{A tree of all of the species with detected bacteriocins.  The five inner rings show gene abundances for immunity, modifier, regulator, toxin and transport genes.  The outer ring shows the total number of bacteriocin-associated gene blocks detected for each bacterial species. This information is available in tabular form in the Supplementary Material.}}	
\label{fig:tree}
\end{figure}
\clearpage




\section*{Tables}
\begin{table}[h!]
\centering
\begin{tabular}{l|r}
Item & Quantity \\\hline
BAGEL Toxins & 479 \\
Gene blocks predicted by BOA with toxin genes & 1,003 \\
Gene blocks predicted by BOA without toxin genes & 83 \\

\end{tabular}

\caption{\label{tab:detected}Number of gene blocks detected by BAGEL and BOA.}
\end{table}
\clearpage

\begin{table}[h!]
\centering
\begin{tabular}{|p{4cm}|c|c|c|c|c|c|c|}
\hline
Name & Genbank ID & \# Gene Blocks & Immunity & Modifier & Regulator & Toxin & Transport\\\hline\hline
\textit{Streptococcus equi subsp. zooepidemicus MGCS10565} & CP001129.1  & 6 & 6 & 8 & 0 & 9 & 11 \\\hline
\textit{Streptomyces griseus subsp. griseus NBRC 13350} & AP009493.1  & 6 & 0 & 0 & 0 & 8 & 14\\\hline
\textit{Leifsonia xyli subsp. xyli str. CTCB07} & AE016822.1	& 5 & 0 & 0 & 0 & 7 & 8\\\hline
\textit{Pusillimonas sp. T7-7} & CP002663.1  & 4 & 1 & 0 & 0 & 6 & 8\\\hline
\textit{Rhodospirillum rubrum F11} & CP003046.1	& 4 & 0 & 0 & 0 & 4 & 4\\\hline
\textit{Streptococcus salivarius CCHSS3} & FR873481.1	& 4 & 5 & 1 & 0 & 4 & 5 \\\hline
\textit{Streptomyces sp. PAMC26508} & CP003990.1	& 4 & 0 & 0 & 0 & 4 & 4\\\hline
\textit{Streptomyces pratensis ATCC 33331} & CP002475.1	& 4 & 0 & 0 & 0 & 4 & 5\\\hline
\textit{Symbiobacterium thermophilum IAM 14863} & AP006840.1	& 4 & 0 & 0 & 1 & 4 & 5\\\hline
\textit{Thermotoga neapolitana DSM 4359} & CP000916.1	& 4 & 5 & 0 & 3 & 4 & 6\\\hline
\end{tabular}
\caption{\label{tab:topdetected}Bacterial species with the highest number of BOA-detected gene blocks}
\end{table}
\clearpage

\begin{table}[h!]
\centering
\begin{tabular}{l|l l l l l | l}
Phylum & Immunity & Modifier & Regulator & Toxin & Transport & \# Genomes\\ \hline
Thermotogae & 0.8 & 0 & 0.2 & 1.4 & 1.8 & 5\\
Proteobacteria & 0.17 & 0.05 & 0.03 & 1.29 & 1.47 & 512\\
Cyanobacteria & 0.13 & 0.13 & 0.0968 & 1.806 & 1.548 & 31\\
Deinococcus-Thermus & 0 & 0 & 0.333 & 2 & 2.667 & 8\\
Euryarchaeota & 0.5 & 0 & 0 & 0.75 & 1 & 6\\
Actinobacteria & 0.337 & 0.045 & 0.067 & 1.708 & 2.168 & 89\\
Spirochaetes & 0.06 & 0.06 & 0 & 1.151 & 1.424 & 33\\
Crenarchaeota & 1 & 0 & 0 & 1 & 1 & 5\\
Firmicutes & 1.182 & 0.578 & 0.399 & 1.594 & 2.059 & 318\\
Bacteroidetes & 0.12 & 0 & 0 & 1.24 & 1.28 & 25\\
\end{tabular}
\caption{\label{tab:phyla_cnts} Mean counts per gene block over all phyla that have more than 5 genomes}
\end{table} 
\clearpage

\begin{table}[h!]
\centering
\begin{tabular}{l|l l l l l | l}
Class & Immunity & Modifier & Regulator & Toxin & Transport & \# Genomes\\ \hline
Spirochaetales & 0.06 & 0.06 & 0 & 1.15 & 1.42 & 33\\
Actinobacteridae & 0.32 & 0.05 & 0.07 & 1.72 & 2.17 & 86\\
Gammaproteobacteria & 0.21 & 0.04 & 0.04 & 1.31 & 1.58 & 244\\
Epsilonproteobacteria & 0.19 & 0 & 0.05 & 1.09 & 1.05 & 21\\
Bacilli & 1.25 & 0.77 & 0.48 & 1.62 & 2.15 & 232\\
Betaproteobacteria & 0.07 & 0.09 & 0.05 & 1.41 & 1.56 & 114\\
Clostridia & 1 & 0.06 & 0.19 & 1.53 & 1.82 & 84\\
Alphaproteobacteria & 0.16 & 0.045 & 0.02 & 1.18 & 1.22 & 110\\
Oscillatoriophycideae & 0.09 & 0.14  & 0.09 & 1.90 & 1.54 & 22\\
\end{tabular}
\caption{\label{tab:class_cnts} Mean counts per gene block over all classes that have more than 10 genomes}
\end{table} 

\clearpage




\section*{Additional Files}
Files are zipped into \textbf{sup\_materials.zip}

\subsection*{00README.txt} the following list.

\subsection*{bagel\_unique.csv} The toxins in the BAGEL database that haven't been identified by BOA. BAGEL ID's literature reference, GenBank cross reference, and protein sequence.

  \subsection*{functions.txt} Per-species counts for each functional ortholog in identified  bacteriocin gene blocks.

  \subsection*{map.txt} A mapping from GenBank ids to gene functions for the detected gene blocks.

  \subsection*{gene\_blocks.fa} The protein sequences of all of the detected gene blocks, their predicted functions and the gene block each protein is in.
  
  \subsection*{lc\_set.txt} Details the literature curated dataset
  \subsection*{lc\_seqs.tgz} Sequence files of the lc\_set genes.
  
  \subsection*{predicted\_gene\_blocks.fa} The protein sequences of all of the detected gene blocks without toxin genes, their predicted functions and the gene block each protein is in.

\end{backmatter}
\end{document}